\documentclass[twocolumn]{aastex7}

\usepackage{longtable}
\usepackage{amsmath}
\usepackage{float}
\usepackage{subcaption}
\usepackage{graphicx}
\usepackage{caption}

\begin{document}
\title{A Correlation Between FRB Dispersion 
Measure and Foreground Large-Scale Structure}

\author[0000-0001-9580-1043]{Maryam Hussaini}
\email{maryam.hussaini@cfa.harvard.edu}
\affiliation{Center for Astrophysics $\mid$ Harvard \& Smithsonian, Cambridge, MA 02138, USA}

\author[0000-0002-7587-6352]{Liam Connor}
\email{liam.connor@cfa.harvard.edu}
\affiliation{Center for Astrophysics $\mid$ Harvard \& Smithsonian, Cambridge, MA 02138, USA}

\author[0000-0001-8235-2939]{Ralf M. Konietzka}
\email{ralf.konietzka@cfa.harvard.edu}
\affiliation{Center for Astrophysics $\mid$ Harvard \& Smithsonian, Cambridge, MA 02138, USA}

\author{Vikram Ravi}
\email{vikram@astro.caltech.edu}
\affiliation{Cahill Center for Astronomy and Astrophysics, MC\,249-17, California Institute of Technology, Pasadena CA 91125, USA}

\author[0000-0001-9855-5781]{Jakob Faber}
\email{jfaber@caltech.edu}
\affiliation{Cahill Center for Astronomy and Astrophysics, MC\,249-17, California Institute of Technology, Pasadena CA 91125, USA}

\author{Kritti Sharma}
\email{kritti@caltech.edu}
\affiliation{Cahill Center for Astronomy and Astrophysics, MC\,249-17, California Institute of Technology, Pasadena CA 91125, USA}
\author{Myles Sherman}
\email{msherman@caltech.edu}
\affiliation{Cahill Center for Astronomy and Astrophysics, MC\,249-17, California Institute of Technology, Pasadena CA 91125, USA}

\begin{abstract}
The distribution of baryons in the Universe remains a fundamental open question in astronomy, and the dispersion measure (DM) of Fast Radio Bursts (FRBs) serves as a valuable tool for probing this cosmic gas. We investigate the impact of the foreground cosmic web on FRB DMs using 61 localized FRBs and public galaxy catalogs. We test for the large-scale structure’s impact on cosmological DM using two methods. First, we searched for a correlation between galaxy number density along the line of sight and extragalactic DM, and found a statistically significant positive correlation ($p$ = $1.76 \times 10^{-5}$). The shape of this correlation contains information about the cosmic baryon distribution, and can also be used to better constrain host galaxy DM by providing an estimate of the cosmic  contribution on a per-source basis. We observe similar correlations in a mock FRB survey based on the IllustrisTNG cosmological simulation, where the DM is dominated by filaments in the IGM and not by halos. Next, we performed a stacking analysis that measures the average excess DM as a function of impact parameter of foreground galaxies to obtain spatial information about how ionized gas is distributed around galaxy halos. We report excess DM in the stacked signal for impact parameters up to Mpc scales ($\sim$\,3\,$\sigma$). Finally, we identified FRBs that do not appear to intersect intervening halos within $r_{vir}$, allowing us to estimate the fraction of baryons that reside in the IGM. We find  $f_{\mathrm{IGM}} \geq 0.69$ at 95\,$\%$ confidence, indicating significant astrophysical feedback.
\end{abstract}

\section{Introduction}
\label{sec:intro}
The distribution of baryonic matter in the Universe has 
implications for multiple critical questions in astrophysics 
and cosmology. Their spatial distribution near and around 
halos of different masses is a fingerprint of different 
feedback scenarios \citep{Dekel, croton, Tumlinson17, Somerville_2015}; measuring this ionized 
gas will answer key questions 
in galaxy formation. In addition to galaxy evolution, 
the unknown baryon powerspectrum 
is a significant nuisance parameter for precision 
cosmology experiments \citep{Castro_2020, euclid, Tr_ster_2022}. Feedback effectively disperses 
matter, leading 
to a smoother Universe with power suppression at non-linear 
scales \citep{amon}. Fortunately, a novel set of probes is 
starting to empirically pin down the distribution 
of cosmic baryons.
\begin{figure*}[t]
    \centering
        \centering
        \includegraphics[width=1\linewidth]{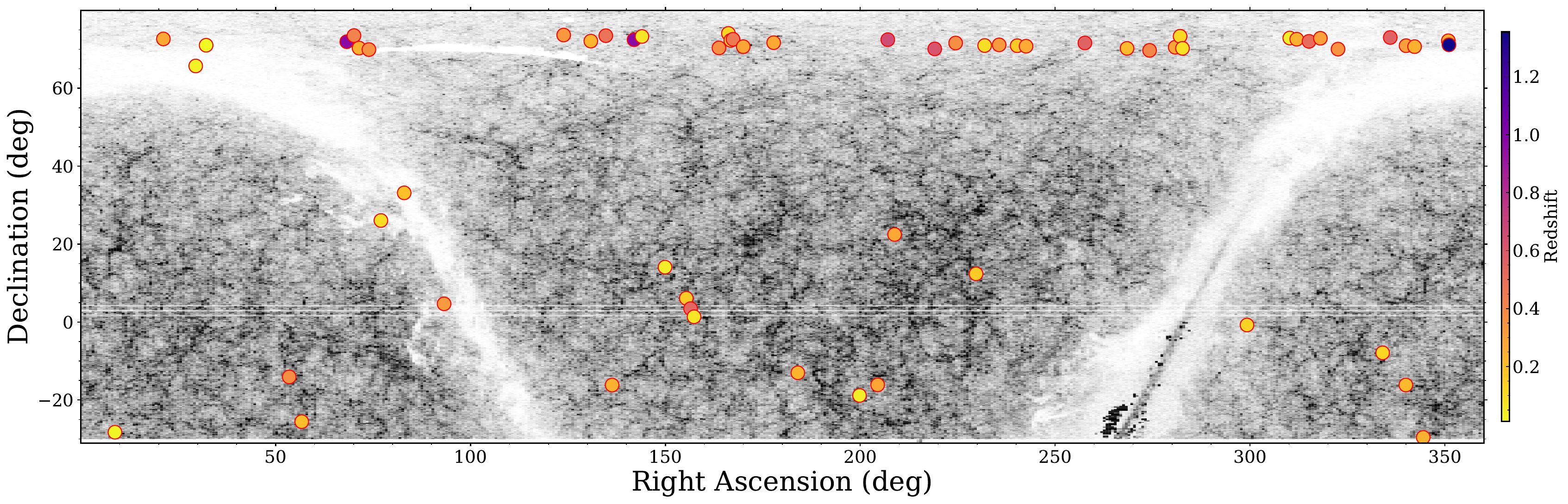} 
    \caption{A 2D histogram showing galaxy number density in the WISE-PS1-STRM for galaxies between $0.10<z<0.20$. The circles mark our FRB sample colored by redshift.} 
    \label{fig:strm-footprint}
\end{figure*}
Fast radio bursts (FRBs) were identified as a powerful tool for cosmology upon their discovery \citep{lorimer2007}, particularly in their ability to measure the missing baryons \citep{fukugita1998, shull2012}. The 
impulsive radio signal undergoes a frequency-dependent dispersion delay due to intervening plasma along the line of sight \citep{petroffreview, cordes-review, mcquinn2014}. This delay is quantified as the dispersion measure (DM), which is the integrated electron number density along the line of sight:

\begin{equation}
\text{DM}(z) = \int_0^z \frac{n_e(z')}{(1+z')^2} \frac{c}{H(z')} dz'.
\end{equation}

The total observed DM for an FRB can be decomposed into contributions from the Milky Way, cosmic gas, and the FRB host galaxy: 

\begin{equation}
    \mathrm{DM}_{\mathrm{obs}} = \mathrm{DM}_{\mathrm{MW}} + \mathrm{DM}_{\mathrm{IGM}} + \mathrm{DM}_{\mathrm{Halo}} + \frac{\mathrm{DM}_{\mathrm{host}}}{1+z_s}
\end{equation}

\noindent In this equation, $\mathrm{DM}_{\mathrm{obs}}$ is the measured (observed) DM; $\mathrm{DM}_{\mathrm{MW}}$ is the contribution from the Milky Way; we define $\mathrm{DM}_{\mathrm{IGM}}$ as dispersion due to ionized matter outside of the virial radius of halos (intergalactic medium; IGM); $\mathrm{DM}_{\mathrm{Halo}}$ is the sum of DM from intervening halos, in this case the observer frame; this includes contribution from the circumgalactic medium (CGM), galaxy groups, and the intracluster medium. $\mathrm{DM}_{\mathrm{host}}$ is the DM contribution of the host galaxy where the FRB originated from; the $(1 + z)$ factor accounts for cosmological time dilation for a source at redshift $z_{\rm s}$. 
We define the cosmological DM, $\mathrm{DM}_{\mathrm{cos}}$, as $
\mathrm{DM}_{\mathrm{cos}} = \mathrm{DM}_{\mathrm{obs}} - \mathrm{DM}_{\mathrm{MW}} - \frac{\mathrm{DM}_{\mathrm{host}}}{1+z_s}.
$

Unlike other probes of the diffuse cosmic baryons, FRB DMs are minimally affected by gas temperature and metallicity, allowing for a direct measure of the total baryon column density out to cosmological distances. 

The positive correlation between extragalactic DM and redshift discovered by \citet{macquart2020} 
demonstrated that FRBs are indeed dispersed by 
cosmic gas. The slope of this correlation, known as the Macquart Relation, is 
proportional to $\Omega_\mathrm{b}\,H_{0}\,f_d$, (where $\Omega_\mathrm{b}$ is the baryon density parameter, ${H}_0$ is the Hubble constant, and $f_d$ is the fraction of baryons in diffuse ionized gas; see Equation \ref{eq:DMcos}) allowing astronomers to statistically detect the bulk of cosmic baryons with a small initial sample of FRB host galaxy redshifts. However, this 
method did not differentiate between gas in the IGM and halo gas, which is central to the missing baryon problem. 

In \citet{connor2024}, the authors used a sample of localized FRBs approximately 10 times larger than that of \citet{macquart2020} to partition cosmic baryons between the IGM, halo gas, stars, and cold gas. They found that 
the statistics of DM and redshift indicate a smooth cosmic gas distribution, with the majority of all baryons residing outside of halos and in the IGM. This fraction, $f_{\rm IGM} \equiv \frac{\Omega_{\rm IGM}}{\Omega_{\mathrm{b}}}$, was found to be roughly three quarters \citep{connor2024}. These results agree with modern cosmological
simulations in which feedback depletes halos of their baryons \citep{illustrisTNG, eagle, simba}. The results are furthermore consistent with other probes of cosmic gas, such as kinematic SZ \citep{hadzhiyska2024} 
% and stacked X-ray observations of galaxies \citep{xx}, 
although 
consensus has not yet been reached on the 
required strength of astrophysical feedback.

Previous works have used FRBs to measure the cosmic baryons without explicitly incorporating foreground galaxy information \citep{macquart2020, yang22, connor2024}. But there is significant information in the distribution of galaxies between the observer and 
the FRB source. A range of methods have been proposed or implemented for incorporating this information, including cross-correlation between FRB positions and galaxy positions \citep{chime-catalog1, Rafiei_Ravandi_2020, Rafiei-Ravandi_2021}; DM/galaxy cross-correlation \citep{shirasaki2017,Madhavacheril2019}; DM/shear cross-correlation \citep{reischke2024calibratingbaryonicfeedbackweak}; and stacking analyses on large samples of unlocalized FRBs \citep{connorravi, WuMcQUinn}. Another method, known as foreground mapping, involves large spectroscopic follow-up campaigns used to reconstruct the 3D cosmic web along each sightline \citep{kglee, flimflamresult, flimflamdr1}, acting as a prior on $f_{\rm IGM}$ (fraction of baryons in the IGM) and $f_{\rm Halo}$ (fraction of baryons contained in the intersected halos). With the release of DESI DR1 \citep{desidr1} and large future spectroscopic galaxy surveys, such methods will only become more popular, especially as the sample of localized FRBs simultaneously grows by orders of magnitude \citep{dsa-2000-whitepaper, chord, masui2025}. 

In this paper, we offer another approach to studying the cosmic baryon distribution using FRBs and the galaxies in their foreground. 
We search for a correlation between DM and the large-scale structure 
using two methods. Both suppose galaxies trace 
dark matter at large physical scales, which also ought to be traced by the cosmic baryons. The first methods comes from the dependence of extragalactic 
DM on the number of 
galaxies in the FRB's direction. A positive correlation is expected because galaxy 
count and cosmic dispersion measure both correlate with the total matter density in the cosmic web. For example, 
a line of sight that travels through multiple voids will pass by fewer galaxies, and will 
also have less DM from the IGM and intervening halos. The exact form of the DM/galaxy density correlation will depend on the distribution of cosmic gas, which can be compared against hydrodynamics simulations or analytic halo models. Independent of our work, \citet{hsu2025} recently found marginal evidence ($p$-value of 0.012--0.032 or $\ \sim \, 2\, \sigma$) for a DM vs. $n_{gal}$ correlation using 
14 localized FRBs. Our work extends those results with a higher significance detection ($\sim 4\,\sigma$) and larger sample size (61 FRBs). 

The second statistic we investigate is excess cosmic DM as a function of impact parameter, stacking across all FRB-galaxy pairs in our sample. This statistic 
offers spatial information on the baryon distribution. To illustrate the limiting cases: If all of the Universe's baryons reside in galaxy halos, then the stacked signal will be proportional to the average electron profile of 
the CGM and most excess DM will be at impact parameters below the virial radius of characteristic halos ($\lesssim250\,$kpc). If halos are, instead, entirely devoid of baryons, then the signal will only come from structure in the IGM, extending 
to Megaparsec scales. Our Universe is somewhere in the middle, but recent evidence points to significant feedback and baryon suppression in halos, leading to a gas-rich IGM \citep{connor2024, hadzhiyska2024}. 
Finally, if the baryons do not trace the dark matter distribution at all or all extragalactic DM comes from the host galaxy, there 
would be no correlation between FRB DMs and foreground galaxy impact parameter. 

The paper is structured as follows:  we first describe our methods (\autoref{sec:methods}) and datasets (\autoref{sec:data}). We compare them with simulations in \autoref{sec:Simulation}. In \autoref{sec:results}, we present the results and correlations found with both statistical methods. In \autoref{sec:discussion} we discuss the implications, limitations, and future prospects of our results. We summarize our findings in the conclusion (\autoref{sec:conclusion}).\\ Throughout 
this paper, we assume the Planck 2018 cosmological parameters \citep{plank2018}.

\begin{figure*}[t]
    \centering

    \begin{subfigure}{\textwidth}
        \centering
        \includegraphics[width=0.98\textwidth]{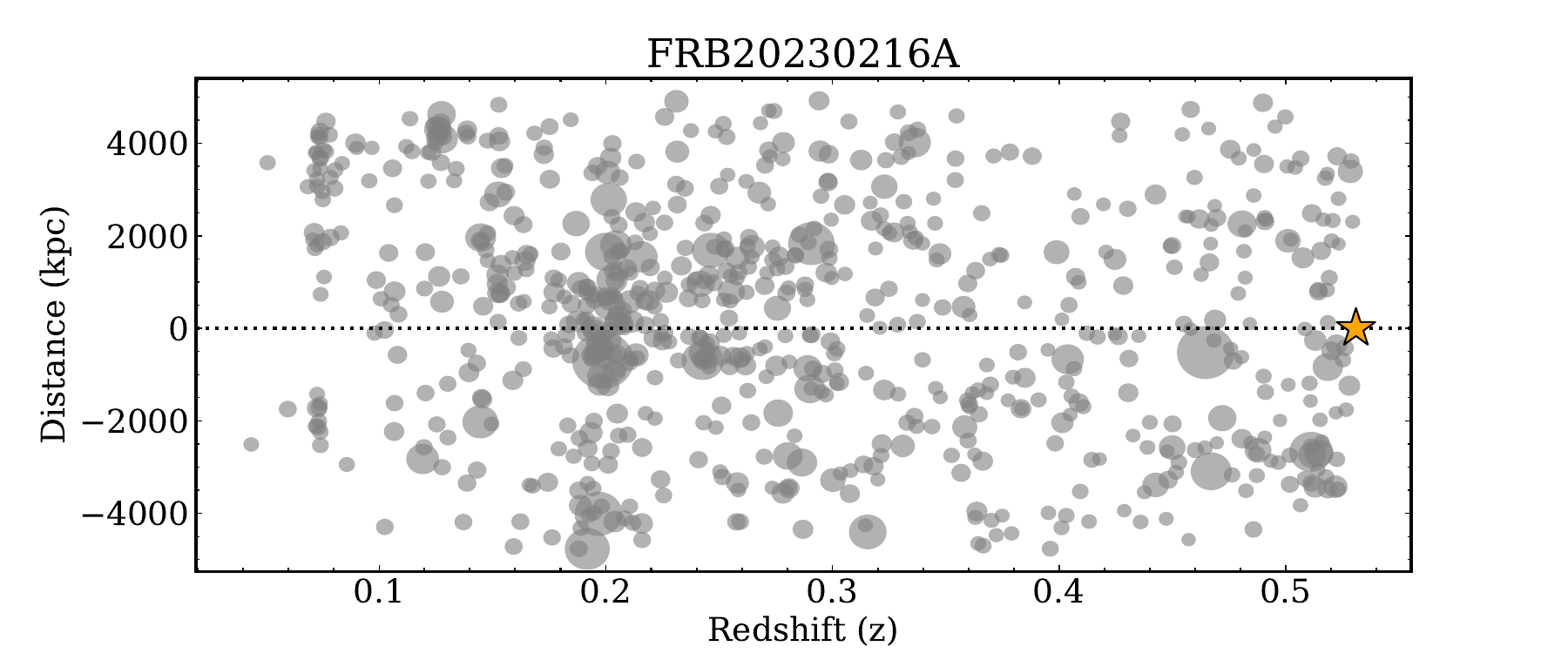}
    \end{subfigure}

    \vspace{-0.5mm}

    \begin{subfigure}{0.83\textwidth}
        \centering
        \includegraphics[trim=0mm 0mm 0mm 0mm,width=\linewidth]{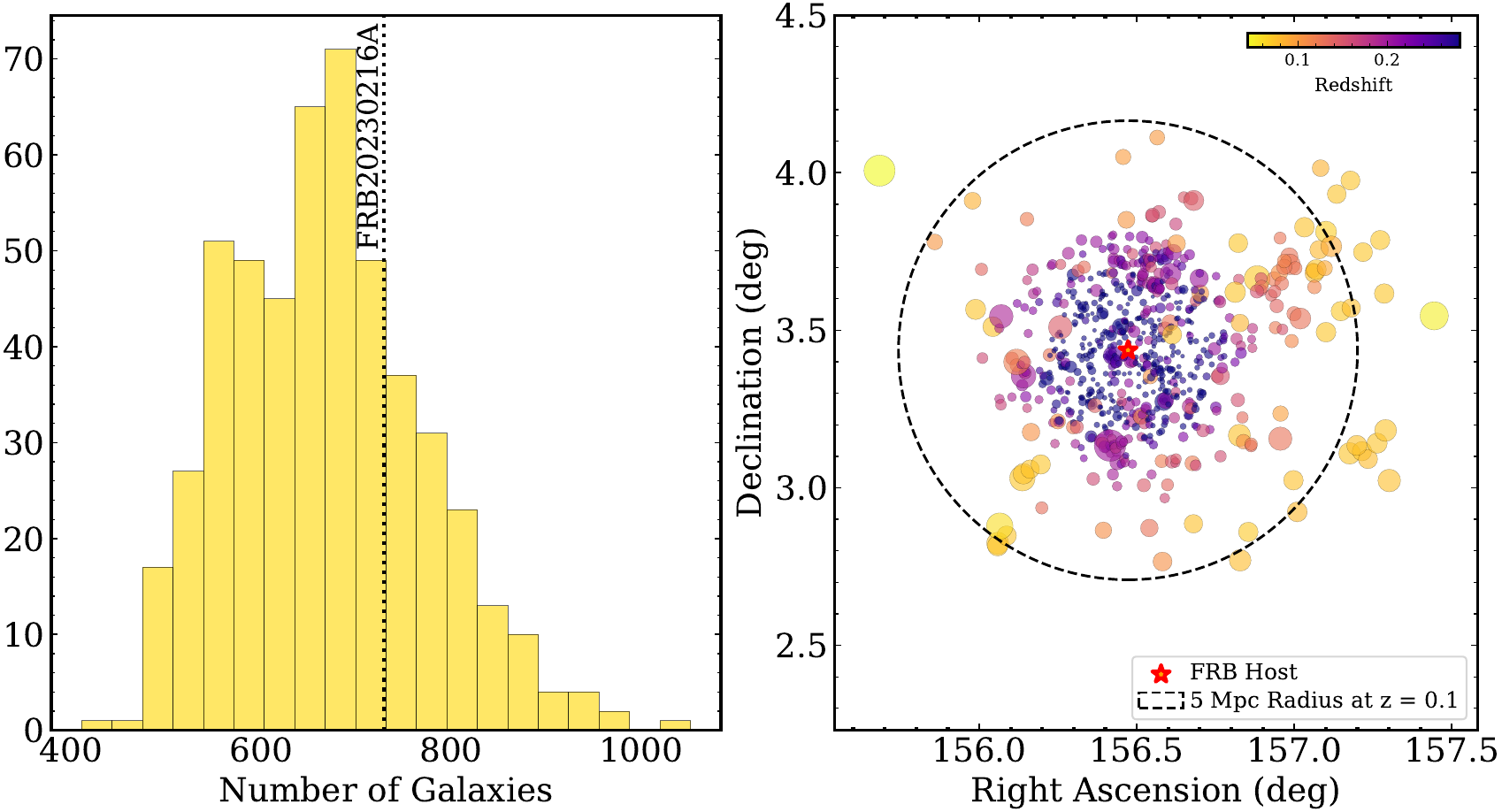}
    \end{subfigure}

    \caption{We use the sightline of DSA-110 discovered FRB\,20230216A to demonstrate our basic approach. This source is at $z=0.53$ and has an extragalactic DM of 788\,pc\,cm$^{-3}$. The \textbf{top panel} shows a side view of foreground galaxies with redshift on the x-axis, and the radius of the cylinder on the y-axis. The FRB source is shown as the gold star at radius = 0 and redshift of 0.53. The grey circles are foreground galaxies from the Legacy survey and their sizes in the vertical direction correspond to their virial radii (scaled to physical kpc). The \textbf{bottom left panel} is a histogram of galaxy counts within 5\,Mpc from 500 random lines of sight at similar galactic latitudes, and to the same redshift as FRB\,20230216A. The \textbf{bottom right} figure shows foreground galaxies within 5\,Mpc of the FRB sightline, colored by their redshift, with FRB\,20230216A at the center as the red star. The black dotted circle shows the angular extent of the 5\,Mpc cylinder at $z=0.1$.}
    \label{fig:frb_combined}
\end{figure*}

\section{Methods}
\label{sec:methods}

We investigated the dependence of baryon column density (inferred from FRB DMs) on the large-scale structure (as traced by galaxies). We first search for a correlation between DM and galaxy number density towards the FRB.
Using the number of galaxies along an FRB's sightline as a proxy for the large-scale structure, we wish to quantify the excess gas (baryonic matter) associated with excess total matter (dark and baryonic matter) in a cylinder towards the FRB. We then investigate how ionized gas is spatially distributed around galaxies by measuring the average excess DM at different impact parameters of foreground galaxies. This `stacking' technique combines data from multiple FRB-galaxy pairs to detect signal that would be too weak to measure in individual sources. Similar techniques have been used in other contexts for decades \citep{Fischer_2000, AndersonCGM, connorravi, WuMcQUinn}. The stacking analysis will allow us to determine not just if there exists a correlation, but which physical scales contribute most. Having mapped the large-scale structure towards each FRB in our sample, we can isolate sightlines that do not appear to intersect the CGM or ICM of foreground halos. An FRB is considered to intersect a halo if its impact parameter is smaller than the halo's estimated virial radius; that is, $b_{\perp} < r_{200}$, where $r_{200}$ is estimated from the galaxy's stellar mass. We use these sightlines that do not cross halos to constrain $f_{\rm IGM}$, the fraction of baryons in the intergalactic medium. 

These are the three methods in our paper: the galaxy number density correlation, a two-point stacking statistic, and a measurement of $f_{\rm IGM}$ from non-intervening sightlines. Our sample is comprised of 61 localized FRBs. For all methods, we analyze the 
 \textit{excess} cosmic DM, $\Delta \text{DM}_{\rm cos}$, subtracting off the Macquart relation (DM$-z$) and the mean host DM for a source at that redshift, $z_s$. 
This is only possible if the 
 FRB has been localized with sufficient angular precision to determine a host galaxy and obtain its spectroscopic redshift. Mathematically, 

\begin{equation}
\Delta \text{DM}_{\text{\rm cos}} = \text{DM}_{\text{obs}} - \text{DM}_{\text{MW}} - \left < \text{DM}_{\text{\rm cos}} \right > - 
\frac{\left < \text{DM}_{\text{host}} \right >}{1+z_s}.
\end{equation}
\noindent Here, Milky Way DM was determined using the \texttt{mwprop.ne2001p} Python library \citep{ocker2024}, calculating the integrated electron density up to a distance of $30\,$kpc in the direction of each FRB. 
$\left <\text{DM}_{\text{host}} \right > $ is taken to be 
150\,pc\,cm$^{-3}$\citep{connor2024, James2022, Khrykin_2024}. Variance in host DM is a key source of noise; covariance of $\text{DM}_{\text{\rm cos}}$ with foreground galaxies is the signal we are interested in.

We calculated the mean cosmological DM contribution, $\langle\text{DM}_{\text{\rm cos}}\rangle$, based on the FRB's redshift:

\begin{equation}
\langle\text{DM}_{\text{\rm cos}}\rangle = f_d\,f_e K \Omega_\mathrm{b} \mathrm{H_0} \int_0^{z_s} \frac{\,(1+z) \, \mathrm{d}z}{\sqrt{\Omega_{\Lambda}+\Omega_\mathrm{m}(1+z)^3}}
\label{eq:DMcos}
\end{equation}

\noindent where $f_d$\footnote{This is technically a function of redshift, but can be taken to be constant for the redshift range of our FRB sample.} is the baryon fraction in the diffuse ionized state, $K = \frac{3c}{8\pi G m_p}$, where $c$ is the speed of light, $G$ is the gravitational constant, and $m_p$ is the proton mass; $\Omega_\mathrm{b}$ is the cosmic baryon abundance, $\Omega_\mathrm{m}$ is the matter density parameter, $\Omega_\Lambda$ is the dark energy parameter, 
$f_e$ is the number of electrons per baryon, taken to be 0.88, and $H_0$ is the Hubble constant. The $f_d$ 
quantity includes both intervening halo gas and the IGM; it 
does not include the host galaxy or the Milky Way contribution. 
We take $f_d=0.93$ from \citet{connor2024}. Our results 
are not sensitive to small changes in this value nor the 
other parameters mentioned here, because 
our statistics do not depend strongly 
on the assumed cosmology. 

\subsection{DM vs. foreground galaxies count}
\label{subsec:ngal}
For each FRB sightline, we counted the number of galaxies within a cylindrical volume extending $5\,$Mpc in radius and spanning up to the FRB's redshift (\autoref{fig:frb_combined}, top panel, lower right panel). The radius was 
chosen to be larger than the several virial radii of the largest halos (massive galaxy clusters) to capture large-scale structure.
To establish a baseline distribution (hereafter ``Control Sample"), we repeated the analysis for 500 random lines of sight extending to the same redshift as that FRB (\autoref{fig:frb_combined}, lower left panel). We ensured the random lines of sight were within $\pm 5^{\circ}$ of the FRB's galactic latitude to maintain similar Milky Way extinction for both the FRB and the control sample. By comparing the galaxy counts between our FRB and the control sample, we quantified the relative  overdensity or underdensity of number of galaxies along each FRB sightline. 

We calculated the z-score to quantify
the extent to which our FRBs' lines of sight are overdense or underdense compared to the control sample, in units of standard deviations:

\begin{equation}
    \text{z-score}_{i} = \frac{n_{gal, i} - \left<n_{gal}\right>}{\sigma_{gal}}.
\end{equation}

Here, $n_{gal, i}$ is the number of galaxies in the direction of the $i^{th}$ FRB within a fixed cylinder of radius $\mathrm{R_{cyl}}$ in proper distance, between $0.01 \leq z_{gal} \leq z_s$; $\left < n_{gal} \right >$ and $\sigma_{gal}$ are the mean and standard deviation of galaxy count in the control sample in the same cylindrical volume, respectively.
In the absence of a well determined physical quantity such as total galaxy mass, the z-score allows us to quantify the galaxy number overdensity or underdensity of the FRB foreground 
across different redshifts. 

We investigated the correlation between the calculated z-scores (representing galaxy number density) and the excess cosmic DM of each FRB. Using the Pearson correlation test, we evaluated the strength of the correlation between these two variables. 
Galaxy data were sourced from WISE-PS1-STRM, and Legacy surveys. See \autoref{sec:data} for more detail. 

A correlation between excess cosmic DM and 
the number of galaxies along an FRB sightline 
would indicate a dependence of 
electron column on the large-scale structure. While this correlation establishes the connection between cosmic gas and structure, future work comparing the correlation's shape and strength with simulation based or analytic models could further constrain the detailed spatial distribution of cosmic gas parameters. 

\subsection{Stacking DM vs. impact parameter}
\label{subsec:stack}
We can also directly search for a spatial dependence 
of DM on impact parameter with foreground galaxies using a classic 
two-point statistic. The procedure analyzes how the cosmic DM contribution varies with the impact parameter ($b_\perp$), which is the projected physical distance between the FRB's line of sight and the center of each galaxy. For each annular bin of impact parameter values, we calculate the mean cosmic DM contribution across all galaxy-FRB pairs falling within that projected distance range. This approach allows us to test for excess dispersion measure as a function of physical separation from galaxies through statistical resampling.
The procedure 
is as follows: 

\begin{enumerate}
    \item Take all galaxy-FRB pairs after the appropriate cuts and compute impact parameter, $b_\perp$.
    \item For each bin in impact parameter, $b_{\perp,i}$ (annuli in projected distance from galaxy center) 
        calculate the mean $\Delta$DM$_{\rm cos}$ over all galaxy-FRB 
        pairs in that annulus. 
    \item Test for statistically significant excess DM as a function of physical separation by comparing to randomized galaxy-FRB pairs and/or jackknife samples.
\end{enumerate}

When analyzing all FRB-galaxy pairs, the same $\Delta$DM$_{\rm cos}$ value from a single FRB contributes to multiple impact parameter bins since each FRB sight line passes by numerous galaxies at different projected distances. This creates inherent correlations between impact parameter bins that must be carefully accounted for in our statistical analysis, as we discuss in \autoref{sec:stacking}.

We focus on Legacy Survey galaxies with photometric redshifts of $0.01 < z_{gal} < 0.5$ and $\log M_* \geq 10$. We limit our sample to 
Legacy in this case because of the relatively accurate photometric redshifts, which are required for computing impact parameters. \citet{zhou2021} also provides stellar masses, allowing for a mass cutoff and virial radius estimation. For the DM-$n_{gal}$ correlation, one only needs to know that the source is in the cylindrical volume, and photo-$z$ errors are 
less important. 
In our case, we also apply a galaxy redshift cut on a per-FRB basis to ensure that all galaxies are in the foreground. The galaxy redshift cut of $z_{gal} \leq 0.8\,z_s$ removes correlations between the FRB host galaxy's relative location in the large-scale structure. For example, if an FRB resides in a galaxy cluster \citep{connor2023, kglee23, chime2025}, it will have excess DM from both the intracluster medium 
and the overdensity in the cosmic web. This would bias our results and undermine our assumption that the sample is homogeneous.
We use only FRBs beyond $z_s=0.10$, allowing the sources to pass by a large number of galaxies, and to reduce the impact of DM$_{\mathrm{host}}$ as a source of noise. 
The fractional impact of host galaxy DM scales as $\sim \frac{1}{z_s\,(1+z_s)}$ for a constant 
rest-frame DM$_{\mathrm{host}}$. We include galaxy-FRB pairs that have an impact parameter smaller than 50\,Mpc and exclude FRBs that are within 2.5\,$\deg$ of the 
edge of the galaxy survey footprint. Finally, we exclude FRB sightlines with 
Galactic latitude less than 5\,$\deg$. 

% The stacked $\Delta \mathrm{DM}_{\rm cos}$ values in impact parameter will be highly correlated across 
% bins, making uncertainty quantification nontrivial. 
Resampling techniques 
can be used to estimate the significance of a detection. We ``scramble'' the FRB positions within the sample itself. The $i^{th}$ FRB takes on the position of the $j^{th}$ FRB, removing any correspondence 
between the assumed foreground galaxy field and measured DM. We also perform a Jackknife test by iteratively removing each FRB sightline and calculating ensemble statistics over the resamplings. We describe these
uncertainty quantification methods in greater detail in Section~\ref{sec:results}.

Finally, we compare our data with simulated FRB data, using the DM ray-tracing catalogs based on IllustrisTNG in \citep{Konietzka}. 
Here, we measure the stacked signal as a function of impact parameter with all intervening halos for a large number of FRB sightlines. 

\section{Data and Sample}
\label{sec:data}

This study examines the correlation between the number density of galaxies along a line of sight and excess cosmic DM for three subsamples of FRB data:

\begin{enumerate}
    \item The full sample, which is comprised of 61 localized FRBs with corresponding WISE-PS1-STRM photometric coverage (\autoref{fig:strm-footprint}).
    \item A subsample of 43 localized FRBs that fall within the Legacy Survey footprint.
    \item 26 FRBs within the Legacy Survey footprint that do not intersect any detectable galactic halos with $\log M_* \geq 10$. An FRB is considered to intersect a halo when its impact parameter is less than the halo’s estimated virial radius, i.e., $b_{\perp} < r_{200}$, where $r_{200}$ is inferred from the galaxy's stellar mass. 
\end{enumerate}
Table~\ref{tab:frb_data} summarizes the properties of our sample.

\subsection{Radio Data}
\label{subsec:radio_data}

There are currently 104 public localized FRBs with candidate host-galaxy redshifts. Of these, 
31 were recently published by CHIME/FRB commissioning its VLBI system with the KKO outrigger \citep{chime2025}. We opt to exclude these from our analysis because of  
selection effects associated with the $2\arcsec\,\times\,60\arcsec$
localization region \citep{james2025}. We note that this limitation will be addressed in future datasets when the full CHIME/VLBI array is operational \citep{masui2025}, which will produce $\sim$\,100 mas 
astrometry.
The majority of our FRB sample comes from two 
radio telescope facilities: Deep Synoptic Array (41 FRBs), 
and Australian Square Kilometer Array Pathfinder (17 FRBs). We describe both of these surveys in detail below. 

\subsubsection{Deep Synoptic Array (DSA-110)}
DSA-110 is a radio interferometer specifically designed for detecting and localizing FRBs with arcsecond precision \citep{Ravi2022Mark, ravi2023b}. The array consists of 96$\times$4.65\,m antennas and is located at Caltech's Owens Valley Radio Observatory (OVRO). DSA-110 operates in the frequency range of 1.28 to 1.53 GHz, with a system temperature of approximately 35 K. The array achieves a typical localization precision of approximately 1-3 arcseconds, depending on the signal to noise ratio of the detection. 41 of our FRBs were discovered by DSA-110 \citep{Ravi2022Mark, law2023_discovery, connor2023,  sherman2023, sharma2024,  connor2024,ravi2019}. 

\subsubsection{Australian Square Kilometer Array Pathfinder (ASKAP)}
ASKAP is a radio telescope array operating as part of the Australia Telescope National Facility \citep{johnston2007, johnston2008, johnston2009, hotan2021}. The facility consists of 36 antennas, each 12 meters in diameter, equipped with phased array feeds capable of dual polarization observations. ASKAP operates across a frequency range of 700-1800 MHz and achieves a field of view of approximately 30 deg$^2$ at 1\,GHz. For FRB detections, ASKAP typically provides localizations with precision of 0.5-10 arcseconds, depending on configuration and signal strength. 17 of our FRBs were detected by ASKAP \citep{bhandari2020, shannon2024, gordon2023}. 

\begin{figure*}[ht]
    \centering
    \subfloat[Full Sample Correlation]{\includegraphics[width=0.45\textwidth]{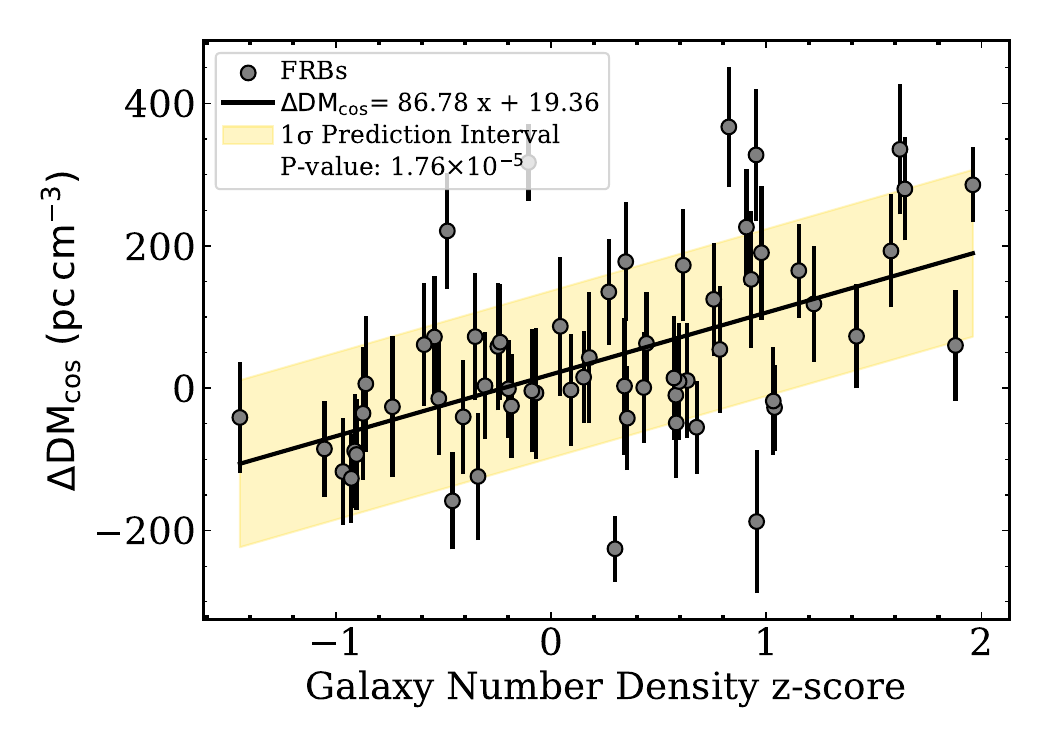}\label{fig:full-corr}}
    \hspace{0.025\textwidth}
    \subfloat[Legacy Sample Correlation]{\includegraphics[width=0.45\textwidth]{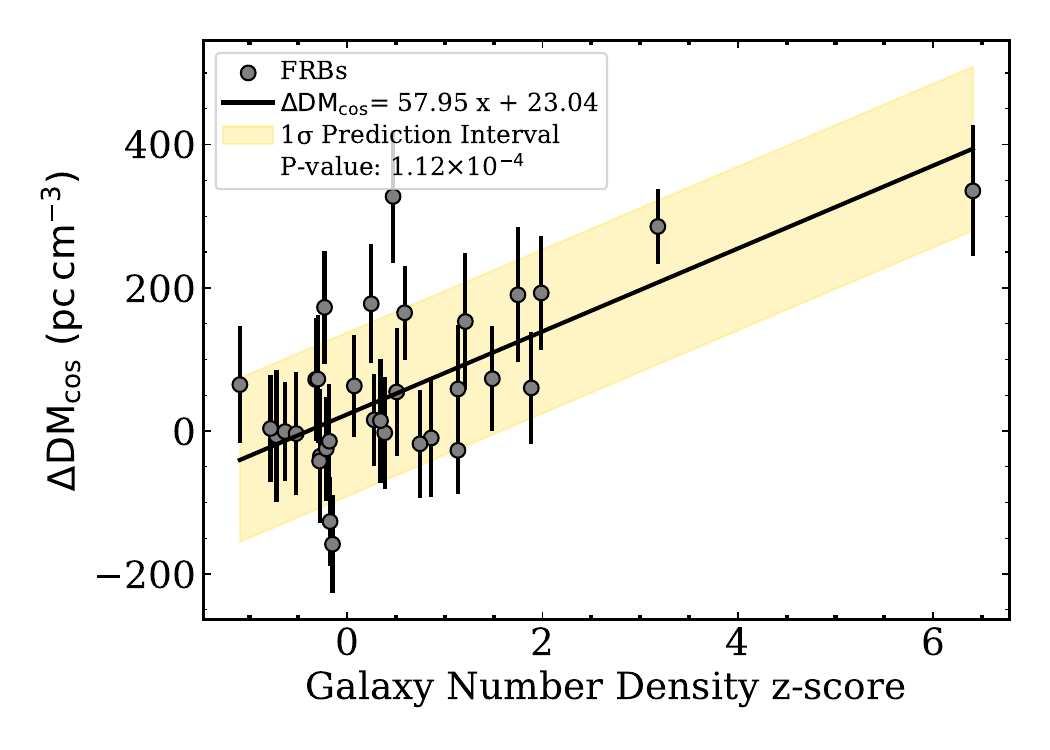}\label{fig:legacy-corr}}
    \caption{Correlation between excess DM and z-score for (a)~the full FRB sample and (b)~the Legacy survey subsample. The x-axis shows the z-score (number of standard deviations from the mean) of galaxy number density, while the y-axis shows the excess DM. Grey dots represent individual FRBs. Vertical error bars are plotted in black. The black lines show the linear best fit and the yellow shaded region is the $1\,\sigma$ prediction interval of the fit. The inset text displays the $p$-value of the correlation and the best-fit equation. Both samples demonstrate a positive correlation between galaxy number density along the line of sight and excess DM.}
    \label{fig:corr}
\end{figure*}

\subsection{Optical and Infrared Data}
\label{subsec:optical_data}

To identify and characterize galaxies along FRB sightlines, we employ two surveys:

\subsubsection{WISE-PS1-STRM Catalog}
The WISE-PS1-STRM catalog combines data from the Wide-field Infrared Survey Explorer (WISE; \citealt{wright2010}) All-Sky Survey and the Panoramic Survey Telescope and Rapid Response System (Pan-STARRS1; \citealt{chambers2016}) 3$\pi$ DR2 source catalog. This dataset covers approximately 3$\pi$ steradians of the sky in five optical to near-infrared bands ($grizy$, covering approximately 400-1000 nm) from Pan-STARRS1 to a 5$\sigma$ point source depth of $r\sim23.3$ mag, and four mid-infrared bands (3.4, 4.6, 12, and 22 $\mu$m) from WISE with point source sensitivities of approximately 54, 71, 730, and 5000 $\mu$Jy, respectively (5$\sigma$). 

The combined catalog has been processed through a machine learning algorithm to provide source classification and photometric redshift estimates \citep{beck2022} with typical redshift uncertainties of $\delta z/(1+z) \approx 0.03$ for galaxies with $r < 20$ mag. 
This catalog enables us to detect galaxies to z $\approx$ 0.5, with optical completeness based on Pan-STARRS1's 5$\sigma$ point-source depth of r$\sim$23.2 mag \citep{chambers2016} and infrared completeness based on WISE's sensitivity limits \citep{wright2010}.
All of our sample (61 FRBs) falls on WISE-PS1-STRM.

\vspace{15mm}

\subsubsection{Legacy Survey}
The DESI Legacy Imaging Survey \citep{dey2019} covers approximately 14,000 square degrees of the extragalactic sky and provides optical and near-infrared photometric catalogs. The survey combines three distinct projects: the Dark Energy Camera Legacy Survey (DECaLS; \citealt{dey2019}), the Beijing-Arizona Sky Survey (BASS; \citealt{zou2017}), and the Mayall z-band Legacy Survey (MzLS; \citealt{dey2019}). Together, these surveys provide $grz$ optical photometry to a 5$\sigma$ point-source depth of $g=24.0$, $r=23.4$, and $z=22.5$ mag, respectively. The Legacy Survey data products also incorporate mid-infrared observations from WISE.

The Legacy Survey achieves a typical seeing-limited resolution of 1.1-1.3 arcseconds and provides reliable star-galaxy separation down to $r \approx 22$ mag. Photometric redshifts are available through the DESI photometric redshift catalog \citep{zhou2021}, with typical uncertainties of $\delta z/(1+z) \approx 0.02-0.03$ for galaxies with $r < 21$ mag. \citet{zhou2021} also provides stellar mass of the galaxies in the survey, which can be used to calculate the virial radius of the galaxies, and impact parameter of FRBs. 43 of our FRBs fall on the legacy footprint, 26 of which do not interact with any halos (i.e. do not come within the virial radius of any galaxy halos). We restrict our DM vs. foreground galaxy count analysis (\autoref{subsec:ngal}) to 33 Legacy FRBs by only including sources at declination $>-30^\circ$. This cutoff addresses the depth discrepancy between the Dark Energy Survey (DES) and the rest of the Legacy Survey. The substantially deeper DES would bias our correlation analysis by detecting more galaxies in its coverage area compared to other survey components, introducing artificial trends in the relationship between galaxy counts and DM measurements. 
%This is not likely to be an issue for our stacking analysis (\autoref{subsec:stack}) where we measure the excess DM as a function of impact parameter, not absolute galaxy count. The shape of this relation is less affected by the galaxy detection thresholds, and our resampling techniques account for any effects of the survey depth differences. 

\begin{figure}
    \centering
    \includegraphics[width=0.98\linewidth]{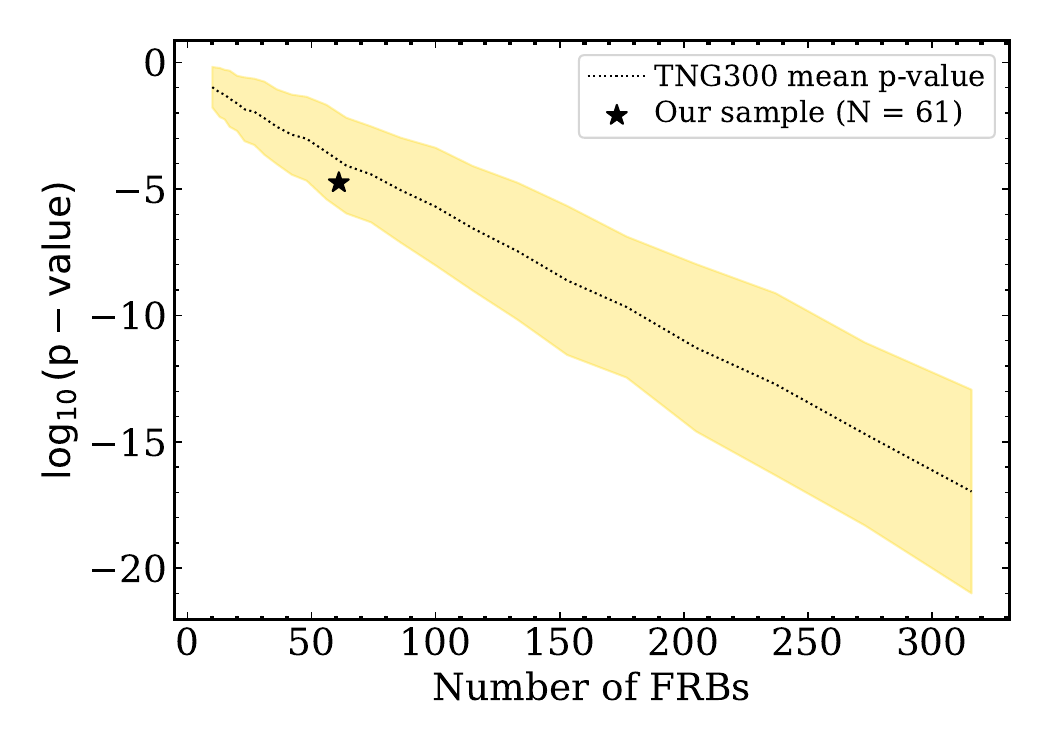}
    \caption{Statistical significance of correlation between excess foreground DM and galaxy number density as a function of sample size. The black star marks our sample size and $p$-value. The dotted line shows the average $p$-value obtained from mock samples using a ray-tracing catalog based on TNG300 from \citep{Konietzka}, and the yellow shaded region is the $90\%$ interval over random drwas of $N_{\rm FRB}$ from the 1000 total sightlines. The y-axis is in $\log_{10}$, and more negative values correspond to stronger correlations. Our result for our sample size of $N_{\rm FRB} = 61$ is consistent with the simulated expectations.}
    \label{fig:tng-p}
\end{figure}

\section{Results}
\label{sec:results}

\subsection{Excess DM vs. galaxy number density}

%We investigated the correlation between galaxy number density along FRB lines of sight and their excess DM using our sample of localized FRBs (Table \ref{tab:frb_data}). 
Our analysis revealed that the number of galaxies along an FRB's line of sight predict excess extragalactic DM.
For our complete sample of public FRBs (N = 61, see Table \ref{tab:frb_data}) we find a positive correlation between galaxy number density z-scores (measure of relative galaxy number density compared to random sightlines of our control sample) and excess DM, with a statistical significance of $p = 1.76 \times 10^{-5}$ ($\sim4.2\, \sigma$; \autoref{fig:full-corr}). The linear regression yields a slope of 87 $\pm$ 12 $\text{pc}\,\text{cm}^{-3}$ per unit z-score increase. When estimating an FRB's redshift from its DM, we can now incorporate cosmic web density information through this relationship. This refinement allows us to account for line of sight variations in baryon distribution beyond the average Macquart relation, potentially improving redshift estimates for localized FRBs by correcting for cosmic web density fluctuations along specific sight lines. As shown in \autoref{fig:tng-p}, our result for a sample size of N = 61 aligns with expectations from simulated DM catalog \citep{Konietzka}. This  suggests the observed DM excess is not attributable to chance alignments with foreground structures.
In \autoref{fig:tng-p}, 
the yellow shaded region indicates the 90$\%$ spread in $p$-values for a fixed number of sightlines within the simulation. For example, with 100 FRBs, the p-value is between $10^{-3.9}$ and $10^{-7.5}$ 90$\%$ of the time. Our result indicates that lines of sight with higher galaxy number densities systematically show higher excess DM values, and fall within the simulated range of $p$-values. Our Legacy Survey subsample (N = 33) shows a correlation with $p = 1.12 \times 10^{-4}$,\ ($\sim 3.8\, \sigma$), shown in \autoref{fig:legacy-corr}.

\subsection{Stacking excess DM vs. impact parameter}
\label{sec:stacking}
\begin{figure*}[t]
    \centering
    \includegraphics[width=0.65\textwidth]{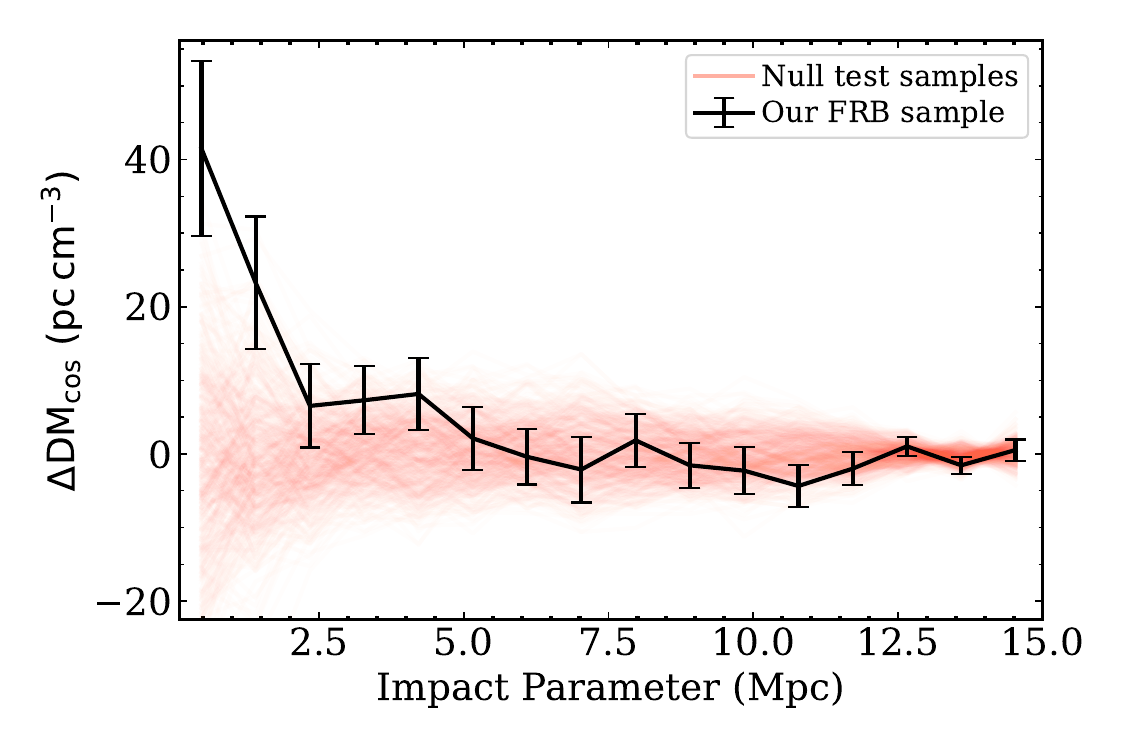}
    \caption{Excess extragalactic DM vs. impact parameter after stacking 
    FRB-galaxy pairs in the Legacy survey footprint. Error bars are estimated via resampling. There is evidence for excess DM up to Mpc scales at $\sim$\,2.5-3.8$\sigma$ significance, 
    depending on the estimator.}
    \label{fig:liamstack}
\end{figure*}

We find evidence for a correlation between excess DM and impact 
parameter in our sample of cosmological FRBs at the $2.5-3.8$\,$\sigma$ level. The smaller value comes from estimating the error via a null test where FRB positions are scrambled. The
$3.8$\,$\sigma$ detection comes from estimating a covariance matrix over jackknife samples. 
The excess DM is a few tens of pc\,cm$^{-3}$. 
The signal extends 
to impact parameters of Mpc scales, exceeding the virial radii of the 
halos on which the FRB DM were stacked. The result constitutes a weak detection, consistent with the DM/$n_{gal}$ correlation we have found at $2-10$\,Mpc scales. If this trend persists with a larger sample, it will be further evidence that DM is not dominated by the 1-halo term of intervening galaxies. We show this result
in \autoref{fig:liamstack}.

\begin{figure}
    \centering
    \includegraphics[width = \linewidth]{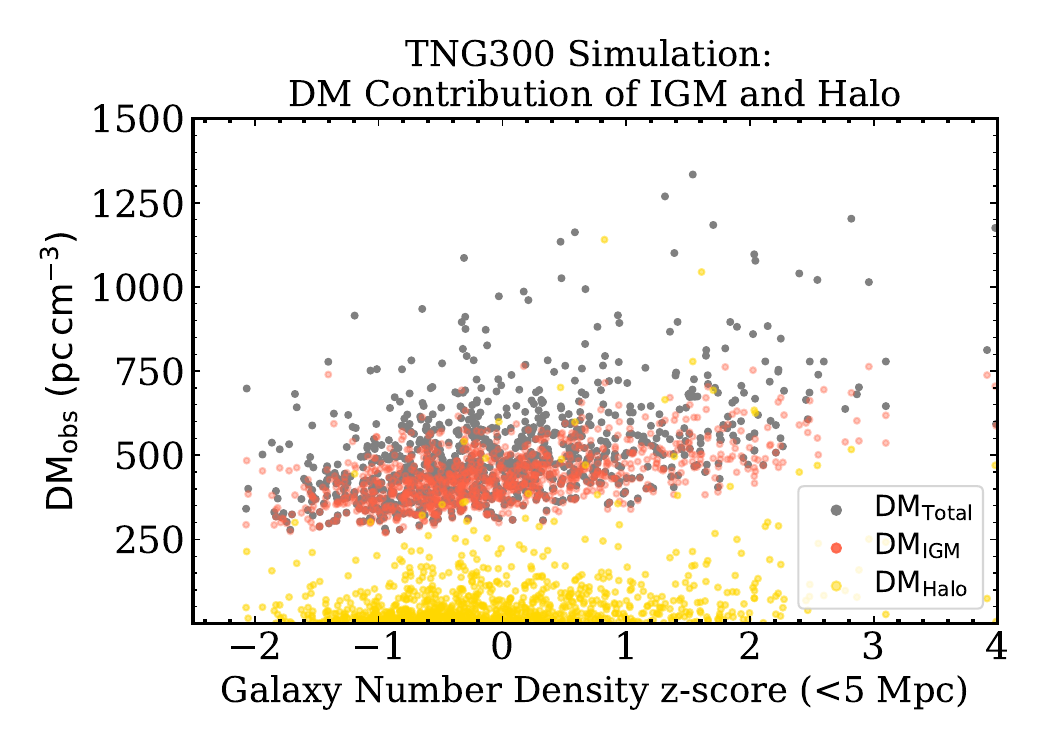}
    \caption{The observed DM coming from IGM and Halos from ray-tracing through the IllustrisTNG simulation \citep{Konietzka} as a function of line of sight galaxy number density. The yellow dots represent the halo contribution (DM from the CGM and ICM), the red dots represent the IGM contribution, and the grey dots are the total DM observed. We observe a similar correlation between excess DM and z-score as we saw in our data. Halos do not seem to contribute to the correlation.}
    \label{fig:sim-corr}
\end{figure}

The use of all FRB/galaxy pairs leads to statistical dependencies 
between impact parameter bins because the same $\Delta$DM$_{\rm cos}$ contributes 
to multiple bins. In some cases, the same galaxy contributes to multiple bins, 
if the FRBs are sufficiently nearby on the sky.
Uncertainty quantification therefore
cannot be a simple Poisson approach related to the square root of 
bin count, nor is the covariance trivial to estimate analytically. 
To solve this problem, we perform a null tests using different resampling techniques.
We first quantify the statistical significance of the stacked signal by computing 
the sum of excess DM below a threshold impact parameter, $b_{\perp, th}$. 
In other words, we calculate the area under the 
$\Delta\,\mathrm{DM}_{\rm cos}$ curve, 

\begin{equation}
    T \equiv \int^{b_{\perp, th}}_{0} \Delta \mathrm{DM}_{\rm cos}(b_{\perp})\,\textup{d}b_\perp
\end{equation}

This number should be zero on average if the baryons and galaxies are 
not correlated and positive if there is excess gas associated with galaxies up to 
$b_{\perp} \leq b_{\perp, th}$. 
We then define our $p$-value 
as the fraction of resampled sightlines with $T \geq T_{data}$. Small $p$-values 
indicate a significant positive excess in cosmological DM correlated with 
foreground galaxy positions. We randomly shuffled the 
FRB positions in our sample, with replacement. 
This should erase any correlation between 
DM and the galaxy field in that FRB's direction. In 500 random shufflings, 
we find that 2 trials result in a larger $T$ than the real data for $b_{\perp, \rm th} = 5\,\rm Mpc$, indicating 
a $p$-value of 0.004. This is roughly at the 2.65\,$\sigma$ level. 

We have also performed a jackknife test by 
iteratively removing each FRB from the dataset 
(with replacement) and computing 
$\Delta \rm DM_{cos}(b_\perp)$. 
If a small number of FRB sightlines dominates the 
apparent excess DM, the variance across jackknife samples will be large. We estimate the significance 
of excess DM from the covariance matrix 
of the jackknife test in the standard way,

\begin{equation}
    \chi^{2} = \boldsymbol{\hat\theta}^{T}\, \mathbf{C}_{\text{jack}}^{-1}\, \boldsymbol{\hat \theta},
\end{equation}

\noindent and 

\begin{equation}
\mathbf{C}_{\text{jack}}
   \;=\;
   \frac{N_{\mathrm{FRB}}-1}{N_{\mathrm{FRB}}}
   \sum_{k=1}^{N_{\mathrm{FRB}}}
   \bigl(\boldsymbol{\theta}_{(k)}-\boldsymbol{\hat\theta}\bigr)
   \bigl(\boldsymbol{\theta}_{(k)}-\boldsymbol{\hat\theta}\bigr)^{\!T}
\end{equation}

\noindent where $\boldsymbol{\hat\theta}$ is the full-sample data vector, in this case $\Delta\rm DM_{cos}$ for $n_{bin}=16$ impact parameter bins, and $\boldsymbol{\theta}_{(k)}$ is the $k^{th}$ jackknife sample of the full dataset. $\mathbf{C}_{\text{jack}}$ is a $n_{bin}\times n_{bin}$ matrix describing the covariance between impact parameter bins. The jackknife covariance results in a $\chi^2=46.5$ 
for 16 degrees of freedom, corresponding to a $p$-value of $8\times10^{-5}$. We therefore reject the null hypothesis of zero excess DM at the $3.8\sigma$ level. 

\begin{figure*}[t]
    \centering
    \includegraphics[width=0.98\textwidth]{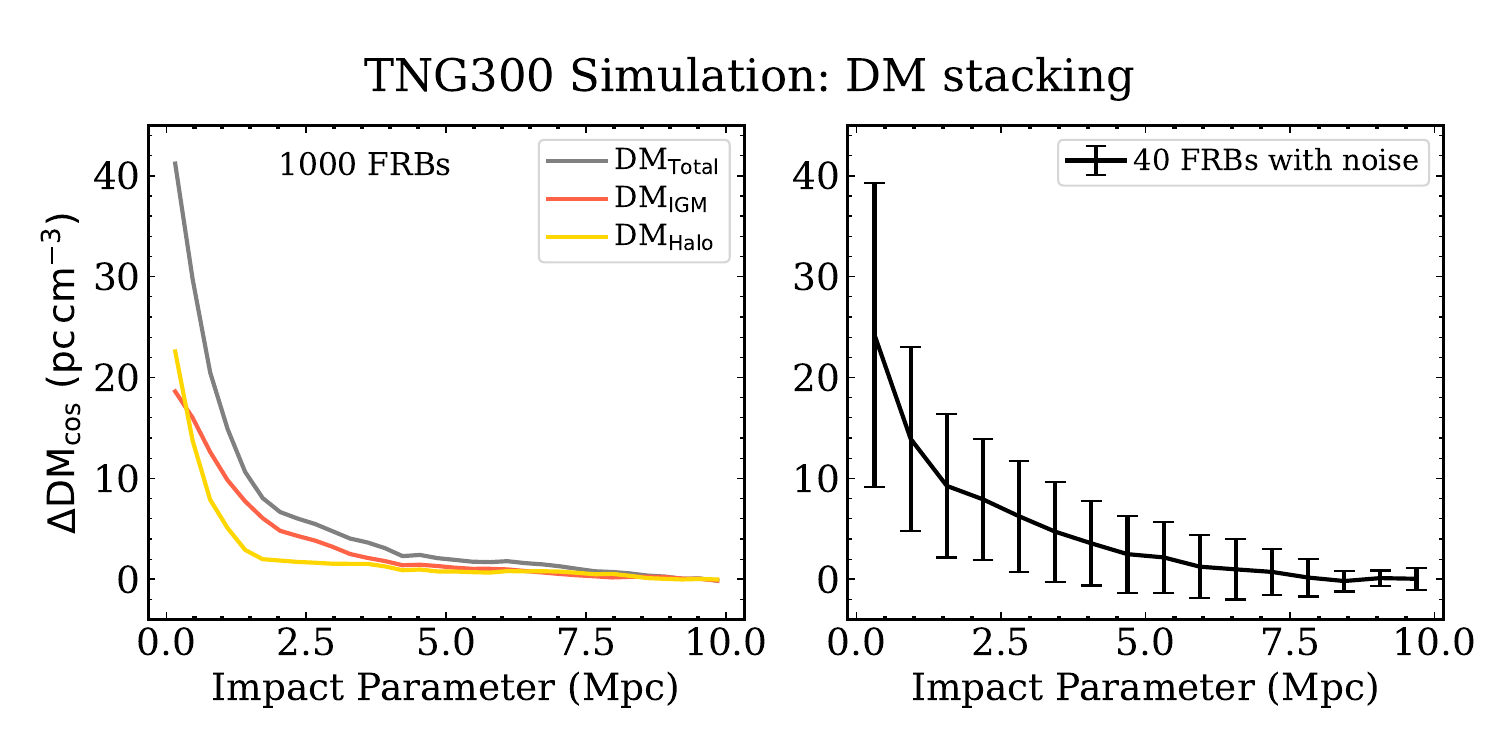}
    \caption{The \textbf{left panel} shows stacked excess DM signal for 1000 simulated FRB sightlines from $z=0.30$ in IllustrisTNG \citep{Konietzka}, separated into total cosmic DM (grey), IGM DM (red), and halo DM (gold). The IGM dominates the DM contribution at all impact parameters $\gtrsim500 \,$kpc, corresponding to 1--2 virial radii of typical halos. In the \textbf{right figure} we show a forward model meant to mimic the observations in this paper, with random redshifts between 0.10 and 0.50, a host DMs drawn from a log-Normal distribution, and noise added for the Milky Way DM subtraction. Even with $\sim$\,40 events, we expect to see excess DM at lower impact parameters at the 2--5\,$\sigma$ level. }
    \label{fig:liamTNG}
\end{figure*}

We compared results to the simulated stacking analysis, described in Section~\ref{sec:Simulation}, where we explicitly separate the contribution 
from halos and the IGM in IllustrisTNG. In Figure~\ref{fig:liamTNG}, we 
show the mean DMs stacked on galaxies in TNG300-1 for 1000 sightlines out from $z=0.30$ and 40 sightlines with random redshifts and added noise. 
We note a tentative similarity between our results (Figure~\ref{fig:liamstack}) 
and the right panel of Figure~\ref{fig:liamTNG}, both in terms of the 
error bars derived by resampling as well as the rough amplitude and extent of 
the excess DM (a few tens of pc\,cm$^{-3}$ to a few Mpc). Large, future 
FRB samples should be compared against a range of simulations 
with varying feedback and cosmological parameters to make strong statements 
about the origin of excess gas associated with halos. 

Previous works have searched for excess DM near foreground galaxies in a sample of roughly 500 unlocalized CHIME/FRB sources \citep{connorravi, WuMcQUinn}. They found marginal evidence 
for higher DMs in FRBs that pass through foreground halos compared with those that do not.
This is difficult with CHIME sources because arcminute positional 
uncertainty precludes the identification of 
intersections with galaxies farther than redshift 0.0125 or $\sim$\,50\,Mpc (i.e. the distance where 
the angular size of a galaxy's CGM is comparable to the 
CHIME/FRB localization). Additionally, without a host galaxy redshift, one cannot subtract off the dominant DM component, $\left < \rm DM_{\rm cos} \right >$.

\subsection{Simulation}
\label{sec:Simulation}
To provide context for our observations, we analyzed data from the TNG300-1 simulation\citep{illustrisTNG,pillepich2019}. The TNG300-1 simulation, part of the IllustrisTNG suite, is a cosmological magnetohydrodynamical simulation modeling a cubic volume of (300 Mpc)$^3$ with over 10 billion resolution elements. It tracks the evolution of dark matter, gas, stars, black holes, and magnetic fields from shortly after the Big Bang to the present day, making it particularly suitable for studying the distribution of baryons in and between galactic halos.
% https://www.tng-project.org/about/
We use a recent simulated DM catalog \citep{Konietzka} that includes extragalactic DMs computed by ray-tracing through the TNG300-1 box \citep{Konietzka}.
DM is computed for 1000 sightlines and is partitioned into 
an IGM component and a halo component. The halo component is defined as Voronoi cells where the dark matter density is 57 times that of the critical density of the Universe. This is approximately the local density of a dark matter halo at its virial radius ($\sim r_{200}$, where the average density within that radius is 200$\rho_c$). The IGM component is defined as cells with local dark matter density of $\leq$57 times that of the critical density of the Universe \citep{artale2022}. 

We first measure $\rm{DM}_{cos}$ vs. galaxy number 
density correlation found in the real FRB data. We use the subhalo catalog 
for galaxy positions in the simulation, counting halos with 
$10^{9}\,h^{-1}\,\mathrm{M}_{\odot} \leq \mathrm{M}_{500} \leq 10^{13}\,h^{-1}\,\mathrm{M}_{\odot}$. The trend 
between excess DM and the galaxy number density 
statistic for 1000 sightlines is shown in Figure~\ref{fig:sim-corr}. The grey dots show the total DM and the yellow and red dots show the
 halo and IGM contribution, respectively. We highlight two key points. The first is that 
ray-tracing through IllustrisTNG \citep{Konietzka} produces a similar positive trend to what we have seen in our data, both in scatter and slope. The second is that this trend is dominated by 
the contribution from the IGM, and
the correlation between halo DM and number of galaxies along the line of sight 
is subdominant.

We also sought to measure the stacking signal of excess DM as a function 
of impact parameter. 
Again, we use the DMs of 1000 synthetic FRB 
sightlines from TNG300-1 \citep{Konietzka} and 
stack that signal at the position of galaxies 
in the simulation's subhalo catalog. 

In Figure~\ref{fig:liamTNG}, we show two stacked DM signals. The left panel is excess DM vs. impact 
parameter for 1000 FRB sightlines all from $z=0.30$, broken into total DM, IGM DM, and halo DM. The IGM contribution dominates 
at all scales larger than a few hundred kiloparsecs, below which the halo DM is slightly higher. The right panel is a forward modeled mock 
FRB survey of 40 sources with redshifts ranging from 0.10 to 0.50 \citep{Konietzka}. This 
sample is designed to mimic our observations. Each of the forty sources 
has a rest-frame host DM randomly drawn from a log-normal 
distribution with $\mu=4.9$ and $\sigma=0.55$ \citep{connor2024}. 
We also add Gaussian noise with mean 0 and standard deviation of 10\,pc\,cm$^{-3}$ for the uncertainty associated with subtracting 
off the Milky Way DM at high Galactic latitude \citep{ocker2020}. 
The error bars in each impact parameter bin come from sampling 40 
sightlines from the 1000 total FRBs, assigning random redshifts and 
host DMs, and computing variance across those resamplings. We find 
that by using the $T$ statistic described in Section~\ref{sec:methods}, 
just 40 events can lead to a moderate detection of excess gas associated 
with halos.

\subsection{Isolating the IGM Contribution}

To distinguish between DM contributions from CGM and the diffuse IGM, we examined a subset of 15 FRBs beyond redshift 0.15 that do not appear to intersect any galaxy halos along their lines of sight. We define intersection as impact parameter less than the virial radius of the galaxy or cluster, $b_{\perp} < r_{vir}$. $r_{vir}$ is defined as the radius within which the density of the galaxy is 200 times the critical density of the Universe. $r_{vir}$ is calculated by converting the stellar mass provided in the Legacy catalog to halo mass using the relation established in \citet{moster2010}.

For this analysis, we only consider galaxies in the Legacy Survey with stellar masses exceeding $10^{10}\,\mathrm{M}_{\odot}$, as the survey becomes increasingly incomplete for lower mass galaxies at the typical redshifts of our FRB sample \citep{dey2019}. Lower mass galaxies should not contribute significantly to the total DM budget, even if their baryon to dark matter ratio is the same as cosmological average. We also 
exclude sightlines that pass through, or are embedded in,  
galaxy clusters with mass greater than $10^{14}\,\mathrm{M}_\odot$, which can impart significant DM \citep{connor2023, chime2025}. We cross-match our FRB sample with several cluster catalogs, including ROSAT RXGCC \citep{xu2022}, the Planck SZ2 cluster catalog \citep{PSZ2}, and a group and cluster catalog made
from DESI Legacy DR9 \citep{DesiDr9Clust}. The cosmic DM of the remaining FRBs ought to come primarily from the IGM. 

\citet{connor2024} constrain the 
total diffuse baryon content, $f_d$, using the mean cosmic DM from Eq.~\ref{eq:DMcos}. We note that an FRB sightline that does not intersect any halos will have $\rm DM_{\rm cos} \approx DM_{\rm IGM}$. 
Therefore, if one can isolate sightlines 
that do not appear to intersect halos, $f_{\rm IGM}$ can be estimated in a similar way. The derived value will be a lower limit because 
sightlines that do not transit through halos will also have less $\rm DM_{IGM}$ due 
to correlations in the large-scale structure. The mean value of $\frac{\mathrm{DM_{\rm cos}}}{g(z)}$
over a sample of IGM-only FRBs gives an estimate of $f_{\rm IGM}$, where  \( g(z) = f_e K\, \Omega_\mathrm{b} \mathrm{H_0} \int_0^z \frac{(1 + z)\, dz}{\sqrt{\Omega_\Lambda + \Omega_\mathrm{m} (1 + z)^3}} \). 

We estimate confidence limits on $f_{\rm IGM}$ by first calculating 
a Gaussian likelihood over the FRB sample,

\begin{equation}
    \mathcal{L}_i = \frac{1}{\sqrt{2\pi\,\sigma_i^2}} \exp{-\frac{(\rm DM_{IGM, obs} - DM_{IGM, mod})^2}{2\sigma^2_i}},
\end{equation}

\noindent and assuming $\mathcal{L} = \prod \mathcal{L}_i$

\begin{equation*}
    \log\mathcal{L} = -\frac{1}{2} \sum^{N_{FRB}}_{i} \log(2\pi\,\sigma^2_i)\, + \frac{\left ( \mathrm{DM_{IGM, obs} - DM_{IGM, mod}}\right)^2}{\sigma^2_i}.
\end{equation*}

\noindent Here, $\rm DM_{IGM, obs}$ is the data and $\rm DM_{IGM, mod}$ is the modeled DM for a given $f_{\rm IGM}$ and FRB redshift. The width of each 
FRB's likelihood function, $\sigma_i$, is the quadrature sum of multiple uncertainty 
contributions,

\begin{equation}
    \sigma^2_i = \sigma^2_{\rm MW} + \sigma^2_{\rm MWhalo} + \sigma^2_{host} + 
     \sigma^2_{\mathrm{LSS}} + \sigma^2_{X}.
\end{equation}

For the error associated with subtracting off the Milky Way's ISM, 
we take $\sigma_{\rm MW} = 0.1\,\mathrm{DM}_{\rm MW}$ \citep{ocker2020} and $\sigma_{\rm MWhalo} = 15$\,pc\,cm$^{-3}$ for the Galactic halo \citep{Ravi2022Mark, cookfrb2023}. 
The host galaxy uncertainty 
comes from the width of the fitted $\mathrm{DM}_{host}$ distribution 
from \citet{connor2024}, which is roughly $100/(1+z_s)$\,pc\,cm$^{-3}$. The large-scale structure uncertainty, $\sigma_{\rm LSS}$, increases in a redshift dependent way, with 
$\sigma_{\rm LSS} \propto \sqrt{z_s}$ as an approximation. 
Finally, we include an uncertainty term, $\sigma_X$, to account for incompleteness in the galaxy catalog that was used to identify sightlines that do not intersect intervening CGM. We take it to be 10$\%$ of the total cosmic DM.
\begin{figure}
     \centering
     \includegraphics[width=\linewidth]{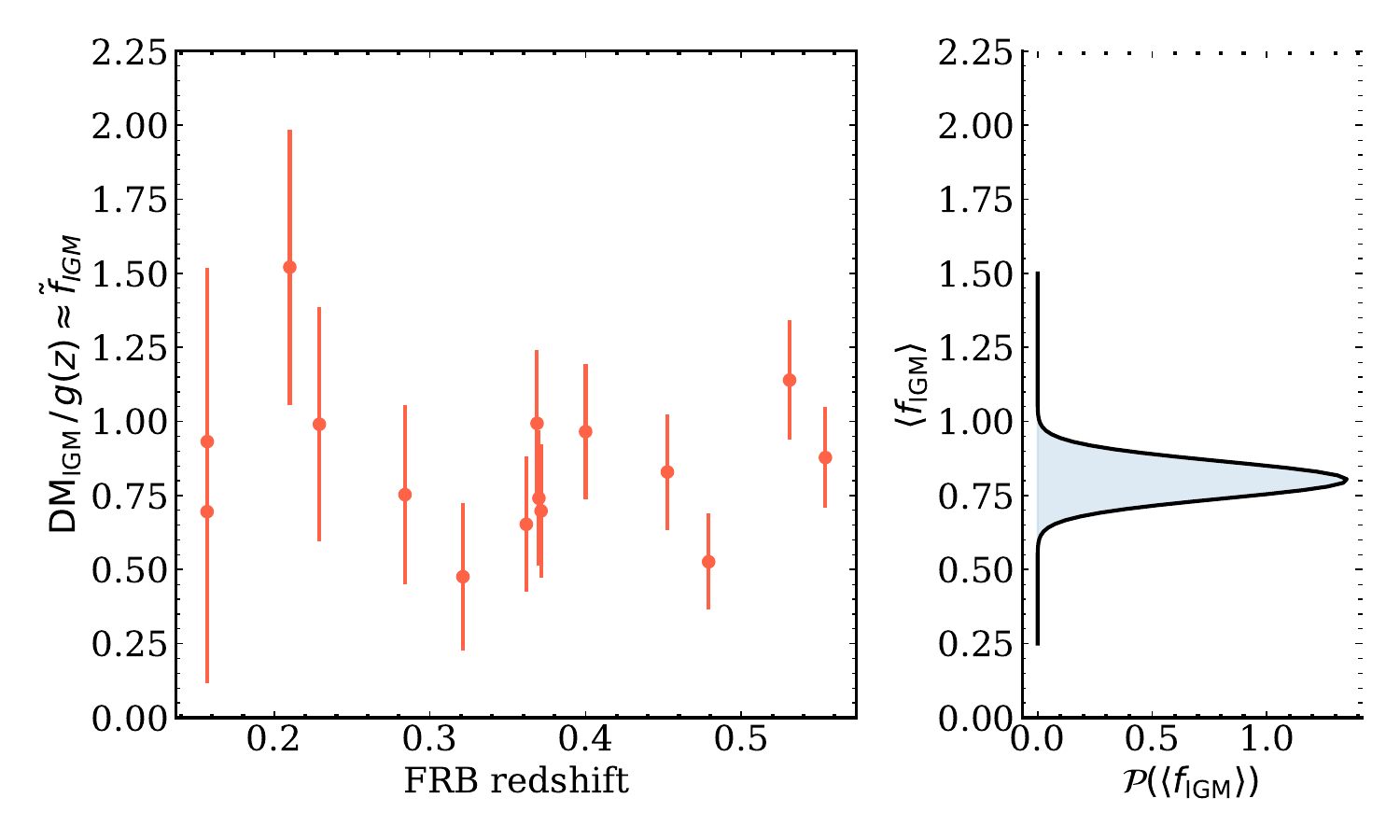}
     \caption{An estimate of the baryon fraction in the IGM 
     based on a sample of FRBs that do not appear to 
     intersect any foreground galaxy halos or clusters. The left panel 
     shows the per-source estimate of $f_{\rm IGM}$ and 
     associated error bars, plotted as a function of redshift. 
     The right panel shows the posterior on the mean of those points, which is a proxy for the IGM baryon fraction, $\mathcal{P}(f_{\rm IGM})$. Our results suggest $f_{\rm IGM}\geq0.69$ at 95$\%$ confidence.
     }
     \label{fig:figm}
\end{figure}

In Figure~\ref{fig:figm} we plot the per-source 
estimates of $f_{\rm IGM}$ as a function of redshift, as well as the resulting 
PDF on the mean, $ \mathcal{P}(\left < f_{\rm IGM} \right >)$, in the right panel. 
Values on the left panel can exceed 1 because of line of sight variations in the IGM and the wide host DM distribution. However, the ensemble average of values is a good proxy for $f_{\rm IGM}$. This distribution gives a lower-bound at 95$\%$ confidence of $f_{\rm IGM} \geq 0.69$. This finding suggests that the IGM is the dominant contributor to total DM, aligning with cosmological simulations 
in which feedback suppresses the baryon content of halos.

The relatively large lower-limit on $f_{\rm IGM}$ 
is consistent with the MCMC fitting method to the full FRB sample in \citet{connor2024}, where it was argued that the steep ``DM Cliff'' 
suggested a minimum DM contribution from the IGM at a given redshift. 
Our finding is not surprising given the redshift distribution of our FRBs and 
the optical depth towards halos. The median redshift of localized FRBs is just under 0.30. Most sightlines to $z=0.30$ do not intersect an intervening CGM or 
galaxy cluster. Thus, most cosmic DM for many of these FRBs is due only to the IGM and it 
is rare for FRBs to have $\frac{\rm DM_{cos}}{z_s} \leq 800$\,pc\,cm$^{-3}$ \citep{connor2024}, which corresponds to 
$f_{\rm IGM}\approx0.80$. We note that our strict definition of a halo intersection as 
less than $r_{200}$ means our data do not preclude the scenario where there is significant gas near and around halos, but outside of the 
virial radius. 

\section{Discussion}
\label{sec:discussion}

We intentionally chose methods that are as model-independent as possible. Other approaches rely on simulation-based inference \citep{connor2024} or a full 3D reconstruction of the 
cosmic web \citep{kglee, Khrykin_2024}. Both approaches have their tradeoffs. For example, the positive correlation between 
$n_{gal}$ and $\Delta$\,DM for a 5\,Mpc aperture does not itself tell us the baryon powerspectrum or the fraction of baryons in halos. It only tells us that the gas traces galaxies on large scales. Similarly, the stacked DM vs. impact parameter signal cannot distinguish between different feedback models without 
a careful comparison against hydrosims or halo models. In future work, we plan to forward model FRB observations in cosmological simulations and investigate both selection effects and the impact of varying 
feedback parameters. Conversely, methods that rely on explicitly modeling 
the cosmic web towards each FRB will be highly informative but sensitive to 
one's model.

\subsection{Correlation Strength and Physical Scale}

The correlation between galaxy counts and excess DM is in part dependent on the radius of the cylindrical volume used to define the line of sight. For our sample of FRBs, we analyzed correlation strength across multiple aperture sizes. At $1000 \, \text{kpc}$, the correlation is weak and not statistically significant ($p = 0.100, \ \sim \, 1.6\, \sigma$). The correlation becomes significant at $3000 \, \text{kpc}$ ($p = 5.16 \times 10^{-5}, \ \sim \, 4\, \sigma$), remains strong at $5000 \, \text{kpc}$ ($p = 1.76 \times 10^{-5}, \ \sim \, 4\, \sigma$), and starts to weaken at $10000 \, \text{kpc}$ ($p = 8.18 \times 10^{-4}, \ \sim \, 3.3\, \sigma$).

The optimal correlation is at approximately 3 to 5 Mpc. This suggests that the excess DM primarily traces intermediate-scale structures as opposed to individual galaxy halos (typically $<1$ Mpc) or large cosmic filaments ($>10$ Mpc). This aligns with theoretical predictions that a significant fraction of cosmic baryons reside in filaments \citep{cen1999, shull2012}.

\subsection{IGM vs. CGM Contributions}

The correlation's persistence in our subsample of FRBs that do not intersect massive galaxy halos strongly suggests that the diffuse IGM, rather than the CGM, dominates the DM contribution. This finding further supports the results of cosmological simulations like TNG300-1 \citep{pillepich2019, illustriscluster}, which predict that a substantial fraction of cosmic baryons reside outside galaxy halos in the diffuse IGM.

Our results provide empirical evidence supporting the theoretical expectation that the majority of baryons exist in a diffuse, warm-hot state between galaxies \citep{dave2001, cen2006}. The correlation between excess DM and galaxy count likely reflects the underlying dark matter distribution, which influences both the galaxy distribution and the content of the IGM.

\subsection{Limitations}

This work has important limitations that affect interpretation and 
stop us from making strong claims about specific astrophysical feedback models. 
Photometric redshift uncertainties introduce scatter in our galaxy position measurements. This is more likely to impact the stacking analysis than the 
galaxy number density correlation, but it adds noise in both cases.
Survey depth limitations affect our ability to detect lower-mass galaxies, especially at higher redshifts. While our analysis accounts for this by focusing on galaxies with stellar masses $>10^{10}\,\mathrm{M}_{\odot}$, the incompleteness could still affect the galaxy count along some sightlines. Variations across the sky due to extinction or galaxy survey incompleteness could lead to systematic uncertainties. 
Finally, we are limited by the sample size of localized FRBs. While growing rapidly, the current sample remains relatively small for robust statistical analysis across multiple subsamples and parameter variations. Host galaxy contributions to DM remain uncertain, which 
again adds noise to our measurements. 
Carefully forward modeled mock FRB and mock galaxy surveys in cosmological simulations could be used to answer these questions and characterize the noise introduced by, e.g., photometric instead of spectroscopic redshifts.

\subsection{Implications for FRB Science}

Our findings provide context for interpreting DM-redshift relationships, 
demonstrating that a significant portion of the scatter in DM for a 
given redshift is due to large structures in the cosmic web. 
This is in contrast to most of the DM scatter being due 
to the host galaxy contribution or the rare intersection of massive halos.

Apart from FRB cosmology, we provide a simple diagnostic (relative galaxy counts in the direction of FRBs) that will allow astronomers to more accurately subtract the cosmic DM contribution on 
a per-FRB basis. With further characterization, this correlation can serve as a calibration tool for future FRB samples. Specifically, to better constrain the host galaxy DM distribution, foreground galaxy catalogs can be used to provide statistical priors for cosmic DM on a source by source basis, allowing for a more accurate DM$_{\rm host}$ measurements. Current methods for estimating $P(\mathrm{DM}_{\rm host})$ typically involve subtracting a model dependent mean DM at the corresponding redshift. The correlation identified in this work provides a more accurate and precise mean DM, as $P(\mathrm{DM}_{\rm cos} | z_s, n_{gal})$ incorporates more information than either  $P(\mathrm{DM}_{\rm cos} | z_s)$ or $\left < \mathrm{DM}(z) \right >$.

\section{Conclusion}
\label{sec:conclusion}
In this work, we studied the relationship between FRB dispersion measures and 
foreground galaxies to study the baryon large-scale structure. 
We summarize our findings 
with the following bullet points:

\begin{itemize}
\item We find a statistically significant positive 
correlation between excess cosmological DM and the number of galaxies within 5\,Mpc along the line of sight of that FRB ($p$-value $\sim 10^{-5}$). The slope and scatter of this correlation is consistent with a mock survey in 
IllustrisTNG \citep{Konietzka}, where the correlation was found to be dominated by contribution from the IGM, as opposed to galaxy halos.

\item We stacked $\Delta$\,DM$_{\rm cos}$ at the positions of foreground galaxies, searching for a dependence of excess DM on impact parameter. We find positive excess at impact parameters up to $\sim$\,Mpc scales at 2.5-3.8$\sigma$ significance, agreeing with the related $n_{gal}$ vs. DM correlation. If the excess out to large scales persists, this will 
again indicate that the cosmic DM is dominated by the IGM or 2-halo term and not the 1-halo term from the CGM.

\item By analyzing the cosmic DM of FRBs that do not appear to intersect galaxy halos or massive clusters, we place a lower limit of $f_{\rm IGM} \geq 0.69$. In other words, at least 69\% of all baryons resides outside of the virial radius of halos, evidenced by the high extragalactic DMs of sources that appear only to be dispersed by the IGM. This finding provides observational support for the theoretical expectation that the majority of cosmic baryons reside in the diffuse gas between galaxies due to astrophysical feedback.

\item Our correlation provides a simple diagnostic to more accurately subtract off cosmic DM, which is useful for studying host galaxies and FRB environments. 

\end{itemize}
\vspace{3mm}

Our results demonstrate the potential of FRBs as probes of the distribution of cosmic baryons after incorporation foreground galaxy information. 
The approaches presented here are intentionally straightforward, 
such as using simple statistics like galaxy number counts. 

As the sample of localized FRBs grows with ongoing surveys from DSA-110, CHIME/FRB Outriggers, and ASKAP, the statistical power of this approach will improve, potentially enabling tomographic mapping of the cosmic web and a 
robust characterization of the baryon power spectrum.
\section{ACKNOWLEDGMENTS}
The DSA-110 is supported by the National Science
Foundation Mid-Scale Innovations Program in Astronomical
Sciences (MSIP) under grant AST-1836018.

The Legacy Surveys consist of three individual and complementary projects: the Dark Energy Camera Legacy Survey (DECaLS; Proposal ID \#2014B-0404; PIs: David Schlegel and Arjun Dey), the Beijing-Arizona Sky Survey (BASS; NOAO Prop. ID \#2015A-0801; PIs: Zhou Xu and Xiaohui Fan), and the Mayall z-band Legacy Survey (MzLS; Prop. ID \#2016A-0453; PI: Arjun Dey). DECaLS, BASS and MzLS together include data obtained, respectively, at the Blanco telescope, Cerro Tololo Inter-American Observatory, NSF’s NOIRLab; the Bok telescope, Steward Observatory, University of Arizona; and the Mayall telescope, Kitt Peak National Observatory, NOIRLab. Pipeline processing and analyses of the data were supported by NOIRLab and the Lawrence Berkeley National Laboratory (LBNL). The Legacy Surveys project is honored to be permitted to conduct astronomical research on Iolkam Du’ag (Kitt Peak), a mountain with particular significance to the Tohono O’odham Nation.

NOIRLab is operated by the Association of Universities for Research in Astronomy (AURA) under a cooperative agreement with the National Science Foundation. LBNL is managed by the Regents of the University of California under contract to the U.S. Department of Energy.

This project used data obtained with the Dark Energy Camera (DECam), which was constructed by the Dark Energy Survey (DES) collaboration. Funding for the DES Projects has been provided by the U.S. Department of Energy, the U.S. National Science Foundation, the Ministry of Science and Education of Spain, the Science and Technology Facilities Council of the United Kingdom, the Higher Education Funding Council for England, the National Center for Supercomputing Applications at the University of Illinois at Urbana-Champaign, the Kavli Institute of Cosmological Physics at the University of Chicago, Center for Cosmology and Astro-Particle Physics at the Ohio State University, the Mitchell Institute for Fundamental Physics and Astronomy at Texas A\&M University, Financiadora de Estudos e Projetos, Fundacao Carlos Chagas Filho de Amparo, Financiadora de Estudos e Projetos, Fundacao Carlos Chagas Filho de Amparo a Pesquisa do Estado do Rio de Janeiro, Conselho Nacional de Desenvolvimento Cientifico e Tecnologico and the Ministerio da Ciencia, Tecnologia e Inovacao, the Deutsche Forschungsgemeinschaft and the Collaborating Institutions in the Dark Energy Survey. The Collaborating Institutions are Argonne National Laboratory, the University of California at Santa Cruz, the University of Cambridge, Centro de Investigaciones Energeticas, Medioambientales y Tecnologicas-Madrid, the University of Chicago, University College London, the DES-Brazil Consortium, the University of Edinburgh, the Eidgenossische Technische Hochschule (ETH) Zurich, Fermi National Accelerator Laboratory, the University of Illinois at Urbana-Champaign, the Institut de Ciencies de l’Espai (IEEC/CSIC), the Institut de Fisica d’Altes Energies, Lawrence Berkeley National Laboratory, the Ludwig Maximilians Universitat Munchen and the associated Excellence Cluster Universe, the University of Michigan, NSF’s NOIRLab, the University of Nottingham, the Ohio State University, the University of Pennsylvania, the University of Portsmouth, SLAC National Accelerator Laboratory, Stanford University, the University of Sussex, and Texas A\&M University.

BASS is a key project of the Telescope Access Program (TAP), which has been funded by the National Astronomical Observatories of China, the Chinese Academy of Sciences (the Strategic Priority Research Program “The Emergence of Cosmological Structures” Grant \# XDB09000000), and the Special Fund for Astronomy from the Ministry of Finance. The BASS is also supported by the External Cooperation Program of Chinese Academy of Sciences (Grant \# 114A11KYSB20160057), and Chinese National Natural Science Foundation (Grant \# 12120101003, \# 11433005).

The Legacy Survey team makes use of data products from the Near-Earth Object Wide-field Infrared Survey Explorer (NEOWISE), which is a project of the Jet Propulsion Laboratory/California Institute of Technology. NEOWISE is funded by the National Aeronautics and Space Administration.

The Legacy Surveys imaging of the DESI footprint is supported by the Director, Office of Science, Office of High Energy Physics of the U.S. Department of Energy under Contract No. DE-AC02-05CH1123, by the National Energy Research Scientific Computing Center, a DOE Office of Science User Facility under the same contract; and by the U.S. National Science Foundation, Division of Astronomical Sciences under Contract No. AST-0950945 to NOAO.
\clearpage
\startlongtable
\begin{deluxetable*}{lccccccp{2in}}
\tablewidth{0pt}
\tablecaption{Fast Radio Burst (FRB) observations with their coordinates, redshifts, dispersion measures, and associated surveys.}
\tablehead{
\colhead{Name} & \colhead{RA} & \colhead{DEC} & \colhead{$z$} & 
\colhead{DM$_{\text{obs}}$} & \colhead{DM$_{\text{exgal}}$} & 
\colhead{Survey} & \colhead{Reference}
}
\startdata
FRB20240210A & 8.7796 & -28.2708 & 0.0237 & 283.73 & 255.10 & ASKAP & \citet{shannon2024} \\
FRB20221116A & 21.2110 & 72.6538 & 0.2764 & 640.60 & 508.27 & DSA & \citet{sharma2024} \\
FRB20180916B & 29.5031 & 65.7168 & 0.0337 & 348.76 & 149.91 & CHIME & \citet{marcote2020} \\
FRB20220319D & 32.1779 & 71.0353 & 0.0111 & 110.95 & -28.83 & DSA & \cite{Ravi2022Mark, law2023_discovery, sherman2023} \\
FRB20200906A & 53.4962 & -14.0832 & 0.3688 & 577.84 & 542.02 & ASKAP & \citet{bhandari2021} \\
FRB20221106A & 56.7048 & -25.5698 & 0.2044 & 343.80 & 309.07 & ASKAP & \citet{shannon2024} \\
FRB20240123A & 68.2625 & 71.9453 & 0.9680 & 1462.00 & 1371.72 & DSA & \citet{connor2024} \\
FRB20240224A & 70.1149 & 73.5128 & 0.4200 & 881.20 & 797.36 & DSA & \citet{law2024} \\
FRB20221113A & 71.4110 & 70.3074 & 0.2505 & 411.40 & 319.73 & DSA & \citet{sharma2024} \\
FRB20220726A & 73.9455 & 69.9295 & 0.3610 & 686.55 & 597.03 & DSA & \citet{sherman2023, sharma2024} \\
FRB20201124A & 77.0146 & 26.0607 & 0.0982 & 411.00 & 271.06 & MeerKAT & \citet{kumar2021} \\
FRB20121102A & 82.9946 & 33.1479 & 0.1927 & 558.10 & 369.70 & VLA & \cite{marcote2017, tendulkar2017} \\
FRB20180301A & 93.2268 & 4.6711 & 0.3304 & 536.00 & 384.39 & ASKAP & \citet{bhandari2021} \\
FRB20231220A & 123.9087 & 73.6599 & 0.3355 & 491.20 & 441.32 & DSA & \citet{connor2024} \\
FRB20221027A & 130.8720 & 72.1009 & 0.2290 & 452.50 & 405.34 & DSA & \citet{sherman2023, sharma2024} \\
FRB20220310F & 134.7205 & 73.4908 & 0.4790 & 462.15 & 415.87 & DSA &\citet{law2023_discovery, sharma2024} \\
FRB20240304A & 136.3308 & -16.1666 & 0.2423 & 652.60 & 578.84 & ASKAP & \citet{shannon2024} \\
FRB20221029A & 141.9635 & 72.4526 & 0.9750 & 1391.05 & 1347.14 & DSA & \citet{sherman2023, sharma2024} \\
FRB20231120A & 143.9840 & 73.2847 & 0.0700 & 438.90 & 395.13 & DSA & \citet{sharma2024} \\
FRB20240201A & 149.9056 & 14.0880 & 0.0427 & 374.50 & 335.93 & ASKAP & \citet{shannon2024} \\
FRB20231226A & 155.3637 & 6.1103 & 0.1569 & 329.90 & 291.83 & ASKAP & \citet{shannon2024} \\
FRB20230216A & 156.4722 & 3.4367 & 0.5310 & 828.00 & 789.52 & DSA & \citet{sharma2024} \\
FRB20211212A & 157.3507 & 1.3605 & 0.0707 & 209.00 & 170.16 & ASKAP & \citet{shannon2024, gordon2023} \\
FRB20220330D & 163.7512 & 70.3507 & 0.3714 & 468.10 & 429.45 & DSA & \citet{sherman2023, sharma2024} \\
FRB20240213A & 166.1683 & 74.0754 & 0.1185 & 357.40 & 317.29 & DSA & \citet{connor2024} \\
FRB20230628A & 166.7867 & 72.2818 & 0.1265 & 345.15 & 306.04 & DSA & \citet{sharma2024} \\
FRB20230712A & 167.3585 & 72.5578 & 0.4525 & 586.96 & 547.76 & DSA & \citet{sharma2024} \\
FRB20240229A & 169.9835 & 70.6762 & 0.2870 & 491.15 & 453.21 & DSA & \citet{connor2024} \\
FRB20230307A & 177.7816 & 71.6951 & 0.2710 & 608.90 & 571.30 & DSA & \citet{sharma2024} \\
FRB20190714A & 183.9797 & -13.0210 & 0.2365 & 504.13 & 465.74 & ASKAP & \citet{heintz2020} \\
FRB20211127I & 199.8087 & -18.8380 & 0.0469 & 234.97 & 192.51 & ASKAP & \citet{shannon2024, gordon2023} \\
FRB20210320C & 204.4587 & -16.1227 & 0.2797 & 384.59 & 345.41 & ASKAP & \citet{shannon2024, gordon2023} \\
FRB20190523A & 207.0650 & 72.4697 & 0.6600 & 760.80 & 723.57 & DSA & \citet{ravi2019} \\
FRB20220105A & 208.8042 & 22.4661 & 0.2785 & 580.00 & 558.06 & ASKAP & \citet{shannon2024, gordon2023} \\
FRB20220418A & 219.1056 & 70.0959 & 0.6220 & 623.45 & 586.76 & DSA & \citet{law2023_discovery, sharma2024} \\
FRB20240119A & 224.4672 & 71.6118 & 0.3700 & 483.10 & 445.19 & DSA & \citet{connor2024} \\
FRB20200430A & 229.7064 & 12.3768 & 0.1610 & 380.00 & 352.92 & ASKAP & \citet{bhandari2021}\\
FRB20230124A & 231.9170 & 70.9681 & 0.0940 & 590.60 & 552.07 & DSA & \citet{sharma2024} \\
FRB20230626A & 235.6296 & 71.1335 & 0.3270 & 451.20 & 412.03 & DSA & \citet{sharma2024} \\
FRB20220920A & 240.2571 & 70.9188 & 0.1585 & 315.00 & 275.11 & DSA &\citet{law2023_discovery, sharma2024} \\
FRB20231123B & 242.5382 & 70.7851 & 0.2625 & 396.70 & 356.45 & DSA & \citet{sharma2024} \\
FRB20221219A & 257.6298 & 71.6268 & 0.5540 & 706.70 & 662.31 & DSA & \citet{sharma2024} \\
FRB20240215A & 268.4413 & 70.2324 & 0.2100 & 549.50 & 501.49 & DSA & \citet{connor2024} \\
FRB20220204A & 274.2263 & 69.7225 & 0.4000 & 612.20 & 561.47 & DSA & \citet{sherman2023, sharma2024} \\
FRB20221012A & 280.7987 & 70.5242 & 0.2840 & 442.20 & 387.84 & DSA &\citet{law2023_discovery, sharma2024} \\
FRB20220914A & 282.0568 & 73.3369 & 0.1138 & 631.05 & 576.36 & DSA & \citet{sharma2023, connor2023, law2023_discovery} \\
FRB20220509G & 282.6700 & 70.2438 & 0.0894 & 269.50 & 213.92 & DSA & \citet{sharma2023, connor2023, sherman2023, law2023_discovery} \\
FRB20210807D & 299.2214 & -0.7624 & 0.1293 & 251.30 & 130.04 & ASKAP & \citet{shannon2024, gordon2023} \\
FRB20220207C & 310.1995 & 72.8823 & 0.0430 & 262.30 & 186.29 & DSA &\citet{law2023_discovery, sharma2024} \\
FRB20220825A & 311.9815 & 72.5849 & 0.2414 & 651.20 & 572.71 & DSA & \citet{sherman2023, sharma2024} \\
FRB20221203A & 315.1295 & 72.0376 & 0.5100 & 602.25 & 518.85 & DSA & \citet{connor2024} \\
FRB20220506D & 318.0448 & 72.8273 & 0.3005 & 396.93 & 312.44 & DSA &\citet{law2023_discovery, sharma2024} \\
FRB20220208A & 322.5751 & 70.0410 & 0.3510 & 437.00 & 335.36 & DSA & \citet{sherman2023, sharma2024} \\
FRB20190608B & 334.0199 & -7.8983 & 0.1178 & 340.05 & 302.86 & ASKAP & \citet{bhandari2020} \\
FRB20230814B & 335.9748 & 73.0259 & 0.5535 & 696.40 & 591.53 & DSA & \citet{connor2024} \\
FRB20210117A & 339.9792 & -16.1515 & 0.2145 & 728.95 & 694.69 & ASKAP & \citet{shannon2024, gordon2023} \\
FRB20230501A & 340.0271 & 70.9221 & 0.3010 & 532.50 & 406.94 & DSA & \citet{sharma2024} \\
FRB20221101B & 342.2158 & 70.6815 & 0.2395 & 490.70 & 359.51 & DSA & \citet{sherman2023, sharma2024} \\
FRB20191228A & 344.4304 & -29.5941 & 0.2430 & 298.00 & 265.15 & ASKAP & \citet{bhandari2021}\\
FRB20220307B & 350.8745 & 72.1924 & 0.2507 & 499.15 & 370.96 & DSA & \citet{law2023_discovery, sharma2024} \\
FRB20230521B & 351.0360 & 71.1380 & 1.3540 & 1342.90 & 1204.15 & DSA & \citet{connor2024} \\
\enddata
\label{tab:frb_data}

\end{deluxetable*}

\bibliography{ref}
\end{document}